\documentclass[aps,
preprintnumbers,superscriptaddress,groupedaddress
]{revtex4}

\usepackage{epsfig,graphicx}
\usepackage{amsmath,amssymb}
\usepackage{bm}

\newcommand{\vslash}{v\hspace*{-5.5pt}\slash}

\def\nslash{n\!\!\!\slash}
\def\bnslash{\bar n\!\!\!\slash}

\newcommand{\beq}{\begin{equation}}
\newcommand{\eeq}{\end{equation}}
\newcommand{\vev}[1]{\langle{#1}\rangle}
\newcommand\pvint{-\mskip-19mu\int}
\newcommand\dy{{\rm d}y\,}
\newcommand\dd{{\rm d}}
\newcommand{\OMIT}[1]{{}}

\begin{document}
\preprint{UCSD/PTH 06-08}


\title{
Shape  and soft functions of HQET and SCET in the 't~Hooft Model}


\author{Benjam\'\i{}n Grinstein}
\affiliation{Department of Physics, University of California at San Diego,
  La Jolla, CA 92093}
\email{bgrinstein@ucsd.edu}
\homepage[]{http://einstein.ucsd.edu/~ben}



\date{\today}

\begin{abstract}
The main application of Heavy Quark Effective Theory (HQET) and of Soft
Collinear Effective Theory (SCET) is in establishing factorization theorems
for exclusive and semi-inclusive decays of heavy mesons. However, the
calculation of the soft factors from the HQET or SCET factorization relations
is, as usual, impeded by the non-perturbative aspect of the strong
interactions. In the hope of gaining some insights into some of these
quantities we compute them in the 't~Hooft model. We find that the
$B$-meson shape function is exactly given by the square of the
$B$-meson light-cone wave-function. The structure of the $B-\pi$
structure function is more complicated: it is given by the product of
wave-functions or by a resonant sum depending on the kinematics. The
result simplifies dramatically in the chiral limit, where it can be
compared with general arguments based on Heavy Meson-Chiral
Perturbation theory. No attempt is made
to use these results for applications to phenomenology.
\end{abstract}



\maketitle



%


\section{Introduction}
One of the most celebrated applications of Heavy Quark Effective
Theory (HQET)
\cite{Isgur:1989vq,Isgur:1989ed,Grinstein:1990mj,Georgi:1990um,Falk:1990yz}
is to the calculation of the rate for the semileptonic decay of a
$B$-meson. It is found that the differential decay rate (with respect
to a single kinematic variable) is systematically predicted as an
expansion in inverse powers of the heavy quark mass, $1/m_b$, with the
leading term completely fixed by symmetry, and hence
calculable\cite{Chay:1990da,Blok:1993va,Manohar:1993qn}. In the case
of charmless semileptonic $B$-meson decays the systematic expansion
breaks down in the ``end-point region,'' namely, when the leptonic
energy is as large as it can be. For
$x \equiv 2E/m_B\approx1-(\Lambda/m_B)^2$ convergence is hopeless, while for
$1-x$ of order unity the expansion is very good. There is a region,
$1-x\approx\Lambda/m_B$, however, for which all orders in the HQET
expansion are equally large. The most singular terms in the expansion
can be formally re-summed into a so-called ``shape-function''
\cite{Bigi:1993ex,Neubert:1993um}

Since the shape function involves the sum of an infinite number of
terms, each characterized by an unknown constant (the matrix element
of a local operator between $B$-meson states), it is largely
unknown. Some basic properties can be readily derived ({\it e.g.},
normalization follows from current conservation), but in practice this is
far from sufficient to determine it with the precision required for
applications. The favored solution to this impasse is to show that the
shape-function is approximately universal and then (approximately)
cancel it from measurements of, say, charmless semileptonic and
radiative $B$-decays.

In this short note we propose to compute the shape function in a the
large $N_c$-limit of QCD in $1+1$ dimensions, the 't~Hooft
model. While the model can not be used to compute parameters of
phenomenological interest, it has been extensively studied in the past
and has given remarkable insights into the dynamics of the strong
interactions. In the context of heavy mesons, calculations in the
't~Hooft model gave some of the earliest indications that the heavy
quark expansion converges rather more quickly for some quantities than
for others: relations among form factors for $B\to D\ell\nu$ and the
predicted normalization of these form factors at zero recoil hold more
accurately than relations among heavy meson decay
constants\cite{Grinstein:1992ub,Burkardt:1991ea}. Some results are
interesting but are unlikely to have a counterpart in four dimensions:
in Ref.~\cite{Grinstein:1994nx} it was shown that in the chiral limit
the form factors for the decay $B\to\pi\ell\nu $ are exactly given by a single
pole formula (and the residue is fixed by symmetry
considerations). And sometimes the results have served to raise
warning flags about possibly unjustified assumptions in the
four-dimensional analysis. For example, in contrast to what has been
argued informally\cite{Bigi:1992su}, a numerical solution to the
model\cite{Grinstein:1997xk} shows that the $1/m_b$ expansion of the
lifetime of the $B$-meson contains corrections of first order in the
expansion parameter, that is, $(1/m_b)^n$ with $n=1$. To be sure, this
result is controversial. It is trivial to show analytically that the
$(1/m_b)^n$, $n=1$ corrections are absent in the chiral
limit\cite{Bigi:1998kc}, but attempts to extend this result to the
non-chiral limit (where the result of Ref.~\cite{Grinstein:1997xk}
applies) are far from rigorous.  Moreover, the numerical result in
\cite{Grinstein:1997xk} is consistent with the theoretical
observations in Ref.~\cite{Grinstein:2001nu,Grinstein:2001zq} that the
$1/m_b$ expansion for the {\it smeared} $B$-width has no $(1/m_b)^1$
term. In sum, the 't~Hooft model is a suitable tool for testing
proposed methods and mechanisms in 4-dimensional QCD that should just
as well apply in 2-dimensional QCD, but it is not likely a good model
for phenomenological applications.

There is another quantity of interest we will compute in this
note. The shape-function can be described as the expectation value in
a $B$-meson state of a non-local b-quark bilinear operator, with a
Wilson line between quarks to ensure gauge independence. A similar
quantity of interest is the matrix element between different states,
say, a $B$-meson and a $\pi$-meson, of a non-local quark bilinear
(with properly chosen flavor quantum numbers). This generalized parton
distribution (GPD) plays a central role in the recently established\cite{GPchiral}
soft pion theorems for certain soft factors in the Soft Collinear
Effective Theory
(SCET)\cite{scet,bfps,bs,bpsfact,mass,Bauer:2002aj,ps1,Beneke:2003pa,
Lange:2003pk,Hill:2004if,CK,BHN,HLPW} description of exclusive $B$
decays to final states containing, possibly, several soft pions. This,
in turn, is useful in extending the applicability of SCET to the
calculation of non-resonant final states, as in, for
example, $B\to K\pi\ell\ell$\cite{Grinstein:2005ud} or $B\to
K\pi\gamma $\cite{Grinstein:2005nu,pol}.

The paper is organized as follows. We begin in  section \ref{review}
with a review of the 't~Hooft model. This is  intended to establish
notation and to review some techniques, but it is not recommended as a
place to learn about the model, which can be best done by going to the
original
sources\cite{'tHooft:1974hx,Callan:1975ps,Einhorn:1976uz,Jaffe:1991ib}.
Section \ref{sec:LCWF} shows that the 't~Hooft wave-fucntions for a
meson are, up
to normalization, the light-cone wave-functions for that meson. This
is an old result, but we use it here as as simple example of the
calculational techniques used in Secs.~\ref{sec:SHAPE}
and~\ref{sec:SOFT} where the shape and soft functions are computed,
respectively. In Sec.~\ref{sec:SOFT} we endeavor to determine the
chiral limit of  the shape functions obtained for general values of
the quark masses. This requires some knowledge of Heavy-Light form
factors, and of how the chiral limit is approached for these form
factors, so we have included a review of those results as the first
sub-section in Sec.~\ref{sec:SOFT}, and we have separated the chiral
limit of the soft functions into a separate sub-section as well. 

Section~\ref{sec:DISC} is the most interesting: we discuss the
interpretation of the results just obtained and compare them to the
literature. We find that the method for computing soft functions
proposed in Ref.~\cite{GPchiral} is not quite correct. Moreover, we
propose a simple (but incomplete) fix. And we find that the shape
function is simply the square of the light-cone wave-function. These
two resulsts are tightly connected in the 't~Hooft model and, as we
remark in the concluding section \ref{sec:CONC} it would be
interesting to determine if this connection persists in 4-dimensional
QCD. This would lead to interesting phenomenology. Finally, some
technical issues, concerning scaling functions, have been left to an appendix.

\section{Review of the model}
\label{review}
This model has been extensively studied and our work relies
on technology pioneered by 't~Hooft\cite{'tHooft:1974hx},
Callan, Coote, and Gross\cite{Callan:1975ps}, and Einhorn\cite{Einhorn:1976uz}.
In these papers the bound state equations were derived; and
it was shown that the scattering amplitudes---and the form factor
in particular---can be written entirely in terms of interactions
among the meson bound states, with no quarks in the spectrum
or in the singularity structure of the amplitudes.

We recall the features of the model which
make it solvable,
and refer the reader to the
original papers for details.
The dynamics are defined by the Lagrangian,
  \beq\label{YMLagrangian}
   {\cal L} =- \frac14 \text{Tr}\, F^{\mu\nu}F_{\mu\nu}
   + \sum_{a} 	\bar\psi_a(\gamma^\mu(i\partial_\mu - g_0 A_\mu)  -m_a)\psi_a,
   \eeq
where $A_\mu$ is an $SU(N_c)$ gauge field, $F_{\mu\nu}$ is its field strength
and $\psi_a$ is a Dirac fermion of mass~$m_a$.
In the large-${N_c}$ limit, the gauge coupling is scaled with ${N_c}$:
\hbox{$g^2 = g_0^2N$} is held fixed as $N_c\to\infty$.
The label $a$ runs over  flavors of quark, with
bare masses $m_a$.

The theory is most conveniently quantized in light-cone gauge.
Because there are no transverse dimensions,
setting~$A_-=0$ eliminates the gluon self-coupling.
It also serves to project gamma matrices onto a single component in
any Feynman graph that has just gluon vertices and ($-$) component
current insertions on fermion lines. In this gauge the gluon
propagator is the inverse of $\partial_-^2$. 
The infrared divergence in the gluon propagator, $i/k_-^2$, is regulated
by taking the principal value at the pole.

Spinors are conveniently split into L and R components,
\beq
\psi_R\equiv P_R\psi={\textstyle\frac12}\gamma_+\gamma_-\psi,\qquad \psi_L\equiv P_L\psi= {\textstyle\frac12}\gamma_-\gamma_+\psi,
\eeq
and in $A_-=0$ gauge $\psi_L$ is not independent,
\beq
\label{psiL}
\psi_{L}=-{\textstyle\frac{im}2\frac{\gamma_-}{\partial_-}}\,\psi_{R}.
\eeq

The leading term of the $1/{N_c}$ expansion is the sum of planar graphs.
The equations for the full
propagator and self-energy can be solved exactly, with an
extremely simple result:
the net effect of all gluons starting and ending on the same
fermion line is just to renormalize the quark mass appearing in the
propagator,
\beq
   m_a^2 \to  \tilde m_a^2 \equiv m_a^2 - g^2/\pi,
\eeq
so the full quark propagator is
\beq
\label{Sdef}
   S^{(a)}(k) = \frac{i k_-}{ k^2 -  \tilde m_a^2  +i\epsilon}.
\eeq
After making this shift, all remaining gluon interactions enter as
ladder-type exchanges.
Crossings would require either gluon self-couplings,
which are absent,
or non-planar graphs, which are higher order in~$1/{N_c}$.
The Yang-Mills coupling constant~$g$ has dimensions of mass, and
we choose units such that $g^2/\pi =1$, leaving
$m_a^2$ as the  dimensionless numbers parametrizing
the theory.
As is well known, there is a discrete spectrum of free mesons.

\begin{figure}
\begin{center}
\includegraphics[height=3cm]{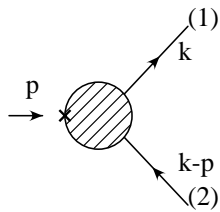}
\end{center}
\caption{\label{fig:Vertex1}
Feynman diagram representation of the full meson-quark vertex  
$S^{(1)}(k)\Gamma^{(12)}(p,k)S^{(2)}(k-p)$ (propagators have been included so the vertex
function is amputated). The
cross denotes an insertion of the current $\bar\psi_1 \gamma_-\psi_2 $. }
\end{figure}

The  full meson-quark vertex, $\Gamma(p,k)$ (see Fig.~\ref{fig:Vertex1}),
satisfies a Bethe-Salpeter-like equation: a gluon exchange between
incoming and outgoing quarks leads to
\beq
\label{eq:Gammaeq}
\Gamma^{(12)}(p,k)-1=-\frac{i}{\pi}\pvint\dd^2q\;\frac1{(q_--k_-)^2}S^{(1)}(q)\Gamma^{(12)}(p,q)S^{(2)}(q-p)~.
\eeq
Here and below, the mark through the integral sign reminds us to take
the principal value of the integrand.  Eq.~\eqref{eq:Gammaeq} shows
that $\Gamma(p,k)$ only depends on $k$ through the variable $x = k_-/p_-$,
so we define $\Gamma(p^2,x)=\Gamma(p,k)$. The equation for the vertex function
is solved in terms of the solutions of the associated eigenvalue
problem, which has the physical interpretation of the bound state
equation:
   \beq\label{BS}
   \mu_n^2 \phi_{n}(x)
   = \left(\frac{ \tilde m_1^2}{x} + \frac{\tilde m_2^2}{1-x}\right)\phi_{n}(x) -
   \pvint_0^1 \frac{\dy}{ (y-x)^2} \phi_{n}(y)
   .
\eeq
The function $\phi_{n}(x)$ is commonly referred to as the 't~Hooft
wave-function of the $n^{th}$ eigenstate, with mass~$\mu_n$,
and~$x=k_-/p_-$ is the fraction of light-cone momentum carried by the
out-going quark. The out-going and in-going quarks may have different
flavor, and hence their masses $m_1$ and $m_2$ are in general
distinct. When needed we will explicitly denote the wave-function by
the flavor of the quarks, as in $\phi^{(12)}_n$.  It is conventional to
choose a standard normalization, $\int_0^1\dd
x\,\phi_n(x)\phi_m(x)=\delta_{nm}$. The $\phi_{n}(x)$ vanish at the
boundaries, and consistency of (\ref{BS}) requires that as $x\to 0$,
$\phi_{n}(x) \sim x^{\beta_1}$, with
\beq\label{Betadef}
   \pi\beta_1\cot\pi\beta_1=1-m_1^2 ~  ,
\eeq
and similarly with the replacements $m_1\to m_2$ and $x\to 1-x$ as $x\to 1$, 
as dictated by the boundary behavior of the
Hilbert transform.  The bound state equation does not have solutions in terms of
known functions, but may be readily solved numerically.

The range of $x$ for the bound states is always in the
interval~[0, 1], and $\phi_n=0$ outside of this range;
but~(\ref{BS}) determines as well the
full meson-fermion-antifermion vertex,
\beq\label{vertexdef}
   \Phi_{n}(z) = \int_0^1 \frac{\dy}{ (y-z)^2} \phi_{n}(y),
\eeq
for values of~$z \notin [0,\,1]$.
This includes~$z$ complex, corresponding to the general case where
one or more of the lines of the meson-quark vertex is not on-shell
in its physical region.
$\Phi_{n}(z)$ is analytic in the complex plane, with a cut on the
real axis from 0 to~1.
When $x \in [0,\,1]$, $\Phi_{n}(x)$ is defined by the
principal value prescription, and
\beq
\label{vertexeq}
   \Phi_{n}(x)
   = \left(-\mu_n^2 + \frac{ \tilde m_1^2}{ x}+\frac{\tilde m_2^2}{ 1-x} \right)
   \phi_{n}(x)
   ,
\eeq
in accordance with~(\ref{BS}).  Since $\phi_{n}(x)$ is finite,
$\Phi_{n}(x)$ has zeros where the first factor on the right vanishes,
and these are the values $x_{\pm}$ where the quarks would be on-shell.
These zeros of the vertex function cancel quark poles in the
propagators of loop amplitudes to ensure that no quark singularities
appear in gauge-invariant Green functions.

All loop integrations are simplified by the fact that $\phi_{n}(x)$ is
a function of $x=k_-/p_-$ only and is independent of $p_+$.  When wave
functions and propagators appear in  loop integrals, only the latter
depend on $p_+$, so the $\int\text{d}p_+$ is over rational functions and can
be computed explicitly by contour integration, leaving a single
integral over one real variable.

The full meson-quark vertex is now easily expressed in terms of the
eigenfunctions,
\beq
\label{Gammagiven}
\Gamma(p^2,x)=1-\sum_n\frac{c_n\Phi_n(x)}{p^2-\mu^2_n}~,
\eeq
where
\beq
\label{eq:cndefd}
c_n=\int_0^1\text{d}x\,\phi_n(x)~.
\eeq
In the interval $0\leq x\leq1$ a handy  alternative expression is available:
\beq
\Gamma(p^2,x)=\left(p^2-\frac{\tilde m_1^2}{x}-\frac{\tilde m_2^2}{1-x}\right)\sum_n \frac{c_n\phi_n(x)}{p^2-\mu^2_n}~.
\eeq

\begin{figure}
\begin{center}
\includegraphics[height=3cm]{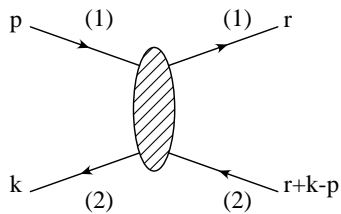}
\end{center}
\caption{\label{fig:scatAmp}
Four quark scattering amplitude.}
\end{figure}

The four quark connected scattering amplitude can be similarly
obtained:
\beq
T=\frac{-ig^2}{2s_-^2}\left(s^2-\frac{\tilde m_1^2}{x}-\frac{\tilde
  m_2^2}{1-x}\right) 
\sum_n\frac{\phi_n(x)\Phi_n^*(z)}{s^2-\mu_n^2}
\eeq
Here the kinematics is as indicated in Fig,~\ref{fig:scatAmp} and we have
defined $s=p-k$, $x=r_-/s_-$ and $z=p_-/s_-$.

Below we will compare some of our results with those obtained from
general considerations using heavy quark and chiral symmetries. It is
therefore useful to present the 't~Hooft wave-functions in the limiting
case of one heavy quark (and a light anti-quark). In heavy quark
effective theory (HQET) Green functions are expanded about the large
momentum of the heavy quark. If the momentum of the heavy meson of
mass $\mu_n$ is $p$, then $v=p/ \mu_n$ is its velocity. We take
$m_1$ to be the heavy quark mass, $m_1\gg 1$ while keeping $m_2$ fixed, 
and write $p=m_1v+\hat p$ and $k=m_1v+\hat k$; see
Fig.~\ref{fig:Vertex1}. Then the argument of the 't~Hooft
wave-function is $x=k_-/p_-=1+(\hat k_--\hat p_-)/m_1v_-+\cdots $ and the large
mass limit is obtained by writing a function of $ \tilde k_-=(1-x)m_1v_-$: 
\beq
\label{eq:HQET-wf}
\psi_n(k_-)\equiv \phi_n(1-k_-/m_1v_-)
\eeq
(we have dropped the tilde since $k_-$ is a dummy variable). Note
that $\hat k_--\hat p_-=k_--p_-$ is the  light quark  
momentum going into the graph. The restriction $x=k_-/p_- \leq1$ is then
the same as the statement that the wave-fucntion is non-vanishing only
if the light quark momentum points out  (when the heavy
meson momentum points in).

Making this change of
variables, writing $\mu_n=m_1+\Lambda_n$ and taking the $m_1\to\infty$ limit, the
't~Hooft wave-function takes the form
\beq
2\Lambda_n\psi(k_-)=\left( \frac{k_-}{v_-}+\frac{\bar
  m_2^2v_-}{k_-}\right)\psi_n(k_-)
-v_-\pvint_0^\infty \frac{\dd q_-}{(q_--k_-)^2}\psi_n(q_-)
\eeq
It is easy to check that this is the equation one obtains starting
directly from the HQET Lagrangian\cite{Burkardt:1991ea}.  

\section{Light-cone wave-function}
\label{sec:LCWF}
Define the light cone wave-functions for a meson $M_n$ by
\beq
\label{lcwfdefd}
\tilde \phi_\pm(\xi)= -\frac{i}{f_n}\int \frac{\text{d}x^-}{2\pi}\,
e^{-i(1-\xi) x^-p_-}\vev{0|W_\pm(x^-)|M_n(p)}
\eeq
where the gauge invariant non-local operators $W_\pm$, defined by
\beq
\label{eq:Wdefd}
W_\pm(x)\equiv\bar \psi(x) P[e^{i\int_o^x A}]\gamma_\pm\gamma_5 \psi(0)
\eeq
have been evaluated along the $x^+=0$ line. In the definition of $W_\pm$, 
$\int\! A=\int\! \dd y^\mu \, A_\mu(y)$, so in our computations, which are
performed in  light cone gauge, this is   $\int\! \dd y^+\,A^-$.
The decay constant  $f_n$ is defined through
\beq
\label{eq:fndefd}
\vev{0|W_\mu(0)|M_n(p)}=\begin{cases}
ip_\mu f_n& \text{if $n$ is even (odd parity)},\\
i\epsilon_{\mu\nu}p^\nu f_n  & \text{if $n$ is odd (even parity)}
\end{cases}
\eeq
The antisymmetric two index tensor $\epsilon_{\mu\nu}$ makes two dimensions
special: it allows us to use an axial current as interpolating field
for scalar mesons (and, similarly, to use a vector current for
pseudo-scalar mesons). It is an easy exercise to compute the
two-current correlator from which it follows that the decay constant
is related to the integral of the wave-function in
Eq.~\eqref{eq:cndefd},
\beq
\label{eq:fngiven}
f_n=\sqrt{\frac{N_c}{4\pi}}c_n
\eeq

The light-cone wave-function $\tilde\phi_-$ is given, up to normalization, by the
't~Hooft wave-function. It is useful to review the derivation of this
fact since the procedure is the same as in the more involved
calculation of other soft functions below.  Consider 
\beq
\label{eq:3point}
\int \dd^2y\;e^{-ip\cdot y} \vev{0|T[W(x^-)\bar\psi\gamma_-\psi(y)]|0}~.
\eeq
In light-cone gauge this is really a three point function. To see this,
take the path in the integral in Eq.~\eqref{eq:Wdefd} to be a
straight line of $y^+=$~constant, from $y^-=0$ to $y^-=x^-$. After 
some spinor algebra, we have
\begin{equation}
\label{split}
\begin{aligned}
W_-(x^-)&=\psi_R^\dagger(x^-)\psi_R^{\phantom{\dagger}}(0)~,\\
W_+(x^-)&=\psi_L^\dagger(x^-)\psi_L^{\phantom{\dagger}}(0)~.
\end{aligned}
\end{equation}
Therefore, the correlator in Eq.~\eqref{eq:3point} is related to the
diagram in Fig.~\ref{fig:Vertex1} by Fourier transformations.  In
order to perform the computation a simple trick proves useful. From
the three point function
\begin{equation}
\label{eq:LCWF-pole}
\Gamma(p,k)=
\int\!\! \dd^2x\;\dd^2y\;e^{-ip\cdot x+i(k-p)\cdot y} 
\vev{0|T[\psi^\dagger_R(y)\psi_R^{\phantom{\dagger}}(0)\psi^\dagger_R\psi_R^{\phantom{\dagger}}(x)]|0}~,
\end{equation}
which is a non-gauge independent quantity given by
Eq.~(\ref{Gammagiven}), determine
\beq
\int \dd^2y\;e^{i(k-p)\cdot y} \vev{0|T[\psi^\dagger_R(y)\psi_R^{\phantom{\dagger}}(0)]|M_n(p)}~,
\eeq
by extracting (and properly normalizing) the residue of the pole for
the $n$-th meson state $|M_n(p)\rangle$. At this point one can
Fourier transform back to $x$ space by integrating over $\int \dd^2k\;
e^{-ik\cdot x}$ and obtain $\vev{0|W(x^-)|M_n(p)}$ by setting
$x^+=0$. This is the gauge invariant integrand of the definition of
the light-cone wave-function in Eq.~(\ref{lcwfdefd}).  The light-cone
wave-function is the  the
Fourier transform of this  with respect to $x^-$ (which undoes one of
the previous Fourier transforms).

Let's see how this works,  explicitly. Extracting the pole,
expression~(\ref{eq:LCWF-pole}) is
\begin{widetext}
\beq
\int \dd^2y\;e^{i(k-p)\cdot y}
\vev{0|T[\psi^\dagger_R(y)\psi_R^{\phantom{\dagger}}(0)]|M_n(p)}=\frac{\sqrt{4\pi N_c}}{p_-}\Phi_n(\xi)
\frac{k_-}{k^2-\tilde m_1^2+i\epsilon }\frac{k_--p_- }{(k-p)^2-\tilde m_2^2+i\epsilon }
\eeq
where $\xi=k_-/p_-$. Following the strategy outlined above, we invert
the Fourier transform and set $y^+=0$, 
\beq
\vev{0|T[\psi^\dagger_R(y^-)\psi_R(0)]|n}=
\int \frac{d^2k}{(2\pi)^2}\;e^{-i(k_--p_-)y^-}\;
\frac{\sqrt{4\pi N_c}}{p_-}\Phi_n(\xi)
\frac{k_-}{k^2-\tilde m_1^2+i\epsilon }\frac{k_--p_- }{(k-p)^2-\tilde m_2^2+i\epsilon }
\eeq
The integral over $k_+$ can be performed explicitly, by appropriately
choosing a semicircle at infinity in the complex $k_+$ plane to close
the contour of integration,
\beq
\vev{0|T[\psi^\dagger_R(y^-)\psi_R^{\phantom{\dagger}}(0)]|n}=i\sqrt{\pi N_c}\int^{p_-}_0\frac{\dd  k_-}{2\pi}
\; e^{-i(k_--p_-)y^-}\Phi_n(\xi)
\frac1{\mu_n^2-\frac{\tilde m_1^2}{\xi}-\frac{\tilde m_2^2}{1-\xi}}
\eeq

\end{widetext}
Since the argument of $\Phi_n$ is in the unit interval we can use
Eq.~(\ref{vertexeq}) to relate it to the 't~Hooft wave-function,
\beq
\vev{0|T[\psi^\dagger_R(y^-)\psi_R^{\phantom{\dagger}}(0)]|n}=
-ip_-\sqrt{\frac{N_c}{4\pi}}\int_0^1\!\!\dd\xi\;e^{i(1-\xi)  p_-y^-} \phi_n(\xi)
\eeq
The coefficient in front is related to the decay constant by
Eq.~\eqref{eq:fngiven}, and the integral over $\xi$ is un-done by the
Fourier transform that defines the light-cone wave function,
Eq.~\eqref{lcwfdefd}. The result is zero for $\xi<0$ or $\xi>1$ and, in
the unit interval,
\beq
\tilde\phi_{n-}(\xi)=\phi_n(\xi)/c_n~.
\eeq
As a fairly trivial check of the calculation, note that
$\int_0^1\dd\xi\;\tilde\phi_n(\xi)=1$. As advertised, up to the
normalization factor,  the light-cone
wave-function is the 't~Hooft wave-function. This, of course, is a
well known fact, but the method exemplified here with this calculation
is precisely what we will use to calculate the shape and soft functions.

The computation of the light-cone wave-function $\tilde\phi_+$ proceeds
in an entirely analogous manner. It follows from 
Eq.~\eqref{psiL} that the result is obtained by including in the
computation above an additional factor of $m_1m_2/4p_{1-}p_{2-}$, where
$p_i$ are the quark momenta. Therefore,
\beq
\tilde\phi_{n+}(\xi)=-\frac{m_1m_2}{2p_-^2\xi(1-\xi)}\phi_n(\xi)/c_n~.
\eeq
Again we can check this by computing its integral. We need the useful relation
\beq
\label{eq:badphiintegral}
\int_0^1\!\dd x \,\phi_n(x)\frac{m_im_2}{x(1-x)}=(-1)^n\mu_n^2c_n,
\eeq
from which it follows that
\beq
\label{eq:intphiplus}
\int_0^1\dd\xi \tilde\phi_{n+}(\xi)=(-1)^{n+1}\frac{\mu_n^2}{2p_-^2}=(-1)^{n+1}\frac{p_+}{p_-}~.
\eeq
On the other hand, the integral of Eq.~\eqref{lcwfdefd} gives
\beq
\int_0^1\dd\xi \tilde\phi_{n+}(\xi)=-\frac{i}{p_-f_n}\vev{0|W_+(0)|M_n(p)}
\eeq
which agrees with Eq.~\eqref{eq:intphiplus} after use of Eq.~\eqref{eq:fndefd}.

\begin{figure}
\begin{center}
\includegraphics[height=3cm]{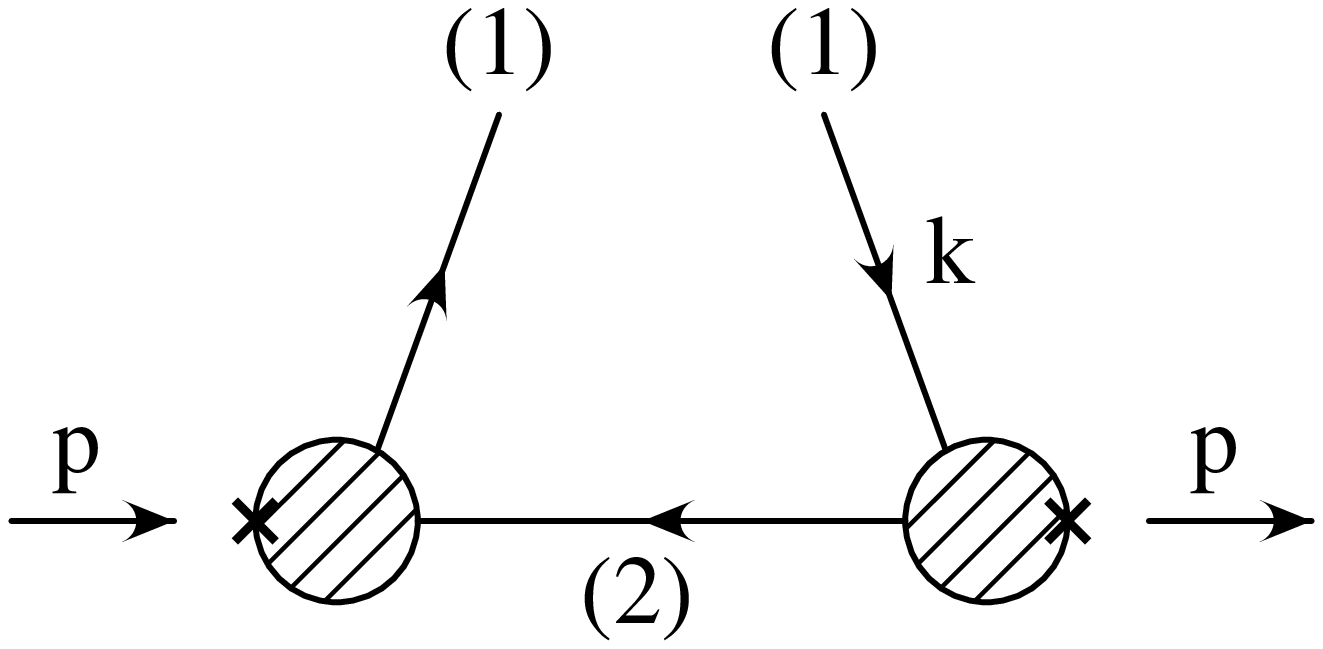}
\hspace{1in}\includegraphics[height=3cm]{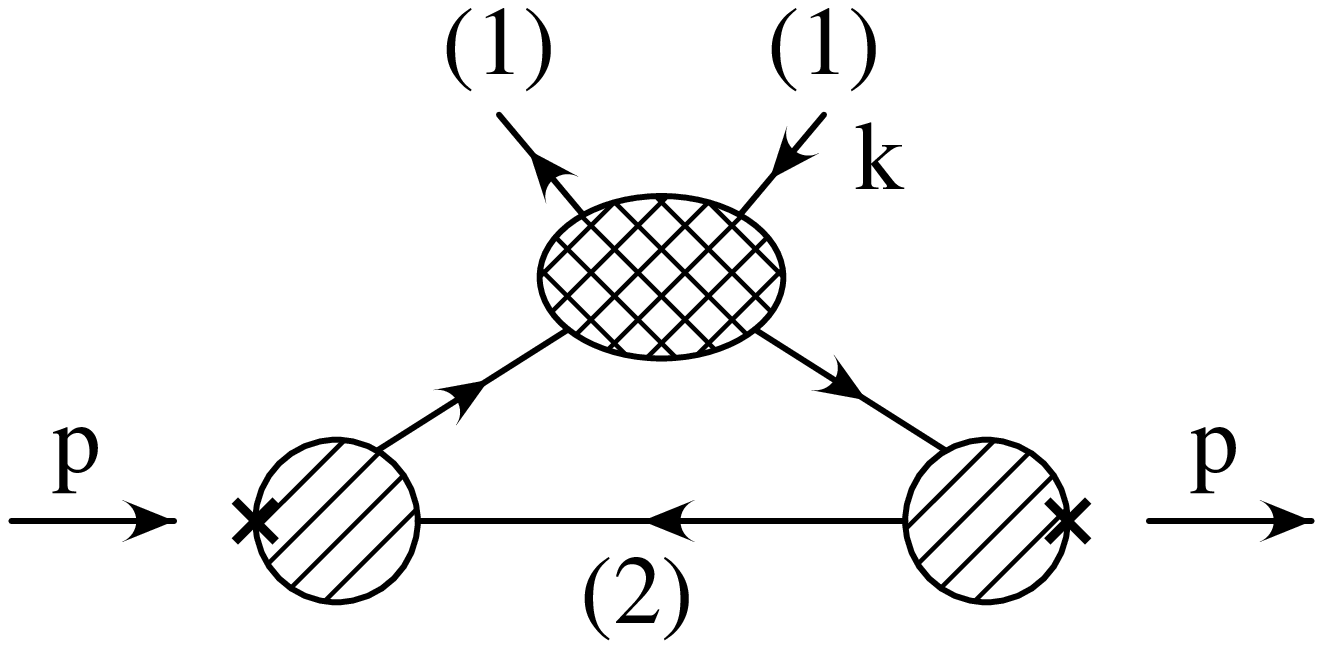}

\end{center}
\caption{\label{fig:shfn1}
Non-resonant (left) and resonant (right) contributions to the Green's function for the soft
function. The crosses denote insertions 
of the $b^\dagger_Ru_R^{\protect\phantom{\dagger}}=\psi_{1R}^\dagger
\psi_{2R}^{\protect\phantom{\dagger}}$ current (or
it's hermitian conjugate), both external lines are $b$-quarks and the
internal line is a $u$-quark.}
\end{figure}

\section{Shape Functions}
\label{sec:SHAPE}
The shape functions are defined by\cite{Bigi:1993ex,Neubert:1993um}
\begin{equation}
f^{(n)}_\Gamma(\xi)  =\frac1{4\pi}\int\dd x^- e^{-i\xi x^- p_-}
\vev{B_n(p)|\bar b(x) P[e^{i\int_0^x\dd A}]\Gamma b(0) |B_n(p)}\Big|_{x^+=0} 
\end{equation}
The notation suggests the meson, $B$, has a heavy $b$-quark and a
light $u$-quark as constituents. In fact, our computation is valid for
arbitrary masses of the quarks, but the notation was adopted since it
is in the context of heavy mesons that shape functions often arise
in practice.

It is convenient to specialize to the case $\Gamma=\gamma_-$. Then, using $\mu=-$
vector-currents to interpolate for the mesons, we are led to consider a
correlator of right-handed quark fields,
\begin{equation}
\label{pre-shfn}
\int\!\!\dd^2x\dd^2y\dd^2z\; e^{ip(x-z)+iky}
\vev{0|T[b^\dagger_R(y)b_R(0)b_R^\dagger(x)u_R(x)u^\dagger_R(z)b_R(z)]|0}~.
\end{equation}
As in the previous section, this quantity can be readily computed but
is not gauge-invariant. However, undoing the Fourier transform over
$y$ and specializing to the line $y^+=0$ does give a gauge independent
quantity.

Two distinct contributions to (\ref{pre-shfn}) are shown in
Fig.~\ref{fig:shfn1}. It is easy to check that
the second graph, involving the full scattering kernel, gives a
vanishing contribution to the shape function. The first graph gives
\beq
N_c\Gamma^{(1\bar 2)}(p,k)\Gamma^{(2\bar 1)}(-p,k-p)S^{(1)}(k)S^{(1)}(k) S^{(2)}(k-p).
\eeq
We have indicated with superscript the quark flavors in these
functions, with $i=1,2$ denoting $q=b,u$, respectively. For the vertex
function the superscript $(1\bar 2)$ indicates an outgoing $b$-quark
and an incoming $u$-quark, and we adopt the same notation for other
related quantities, {\it e.g.}, the 't~Hooft wave-functions.  Applying
LSZ reduction and integrating over $k_+$ this gives
\beq
\phi_n^{(1\bar 2)}(\xi)\phi_n^{(2\bar 1)}(1-\xi)\theta(\xi)\theta(1-\xi),
\eeq
where $\xi=k_-/p_-$. We have kept the subscript $n$ arbitrary,
indicating that the result is more general than we set out to obtain,
namely, it is valid for the shape function of any of the states in the
tower of which the $B$ meson is the ground state (corresponding to
$n=0$). Using $\phi_n^{(2\bar 1)}(1-\xi)=\phi_n^{(1\bar 2)}(\xi)$ we
finally have 
\beq
\label{shfnresult}
f^{(n)}_{-}(\xi)= (\phi_n^{(1\bar 2)}(\xi))^2
\eeq
where it is understood the support is for $0\leq\xi\leq1$ and the subscript in
$f$ reminds us of the choice $\Gamma=\gamma_-$.

From the normalization of the 't~Hooft wave-functions it follows
that
\beq
\int_0^1d\xi f^{(n)}_-(\xi)=1
\eeq 
upon integration of Eq.~\eqref{shfnresult}. This is a minimal test on
our calculations since the normalization of the shape function follows
from charge ($b$-number) conservation.

Finally we express this result in terms of functions in the heavy
quark limit. It is customary to write the shape function in terms of
the residual momentum $\hat k_-$ of the $b$-quark. Our expression for the heavy
quark 't~Hooft wave-function, Eq.~\eqref{eq:HQET-wf}, has the light
quark momentum as argument. To write a heavy quark limit formula for
the shape function as a function of the $b$-quark residual momentum
$\hat k_-$ we use $\Lambda_n=\mu_n-m_1$, so that  $\xi=k_-/p_-=(m_1v_-
+\hat k_-)/(m_1v_- +\Lambda_n v_-)=1+ (k_--\Lambda_n v_-)/m_1v_- +\cdots$. We obtain
\beq
f^{(n)}_-(\hat k_-)=[\psi_n(\Lambda_n-\hat k_-)]^2
\eeq
Note that this has support only for $\hat k_-<\Lambda_n$.

\section{``Heavy-light'' soft function}
\label{sec:SOFT}

\subsection{``Heavy-light'' Form Factors}
In preparation for our study of the soft function we review some basic
results for the transition current form factors. We consider mesons
with different flavor, a $b\bar u$-meson that we refer to, in analogy
with it's four dimensional counterpart, as the ``$B$-meson,'' and a
$u\bar d$ meson, the ``pion.'' Flavors $b, d, u$ will be denoted by a
flavor index $a=1,2,3$, respectively. While we refer to $b$ as the
``heavy'' quark and to $u$ and $d$ as the ``light'' quarks, we do not
make assumptions on the relative sizes of the masses, except when we
consider the chiral limit for which we take $m_2=m_3\to0$ ($m_u=m_d\to0$)
holding $m_1$ ($m_b$) fixed. The form factors $f_\pm$ are defined
by\footnote{It is customary to use the symbol $f$ both for shape
  functions and for form factors. Which quantity is being discussed
  is always clear from context.  }
\beq
\label{ffdefd}
\vev{\pi(p')|\bar d(0) \gamma_\mu  b(0)|B(p)}=(p+p')_\mu f_+(q^2)-(p-p')_\mu f_-(q^2).
\eeq
where $q=p-p'$. Crossing this gives
\beq
\label{ffxdefd}
\vev{0|\bar d(0) \gamma_\mu  b(0)|B(p)\bar \pi(p')}=(p-p')_\mu f_+(q^2)-(p+p')_\mu f_-(q^2).
\eeq
where now $q=p+p'$.

\begin{figure}[t!]
\begin{center}
\includegraphics[height=3cm]{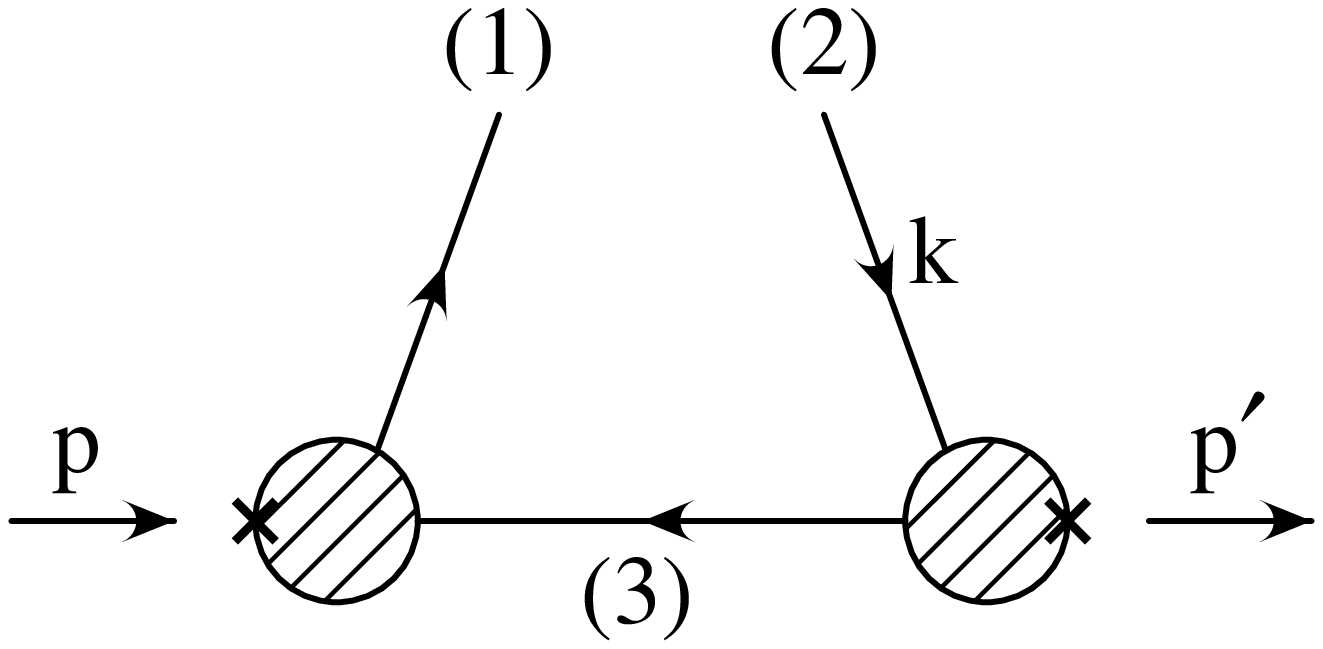}
\hspace{1in}
\includegraphics[height=3cm]{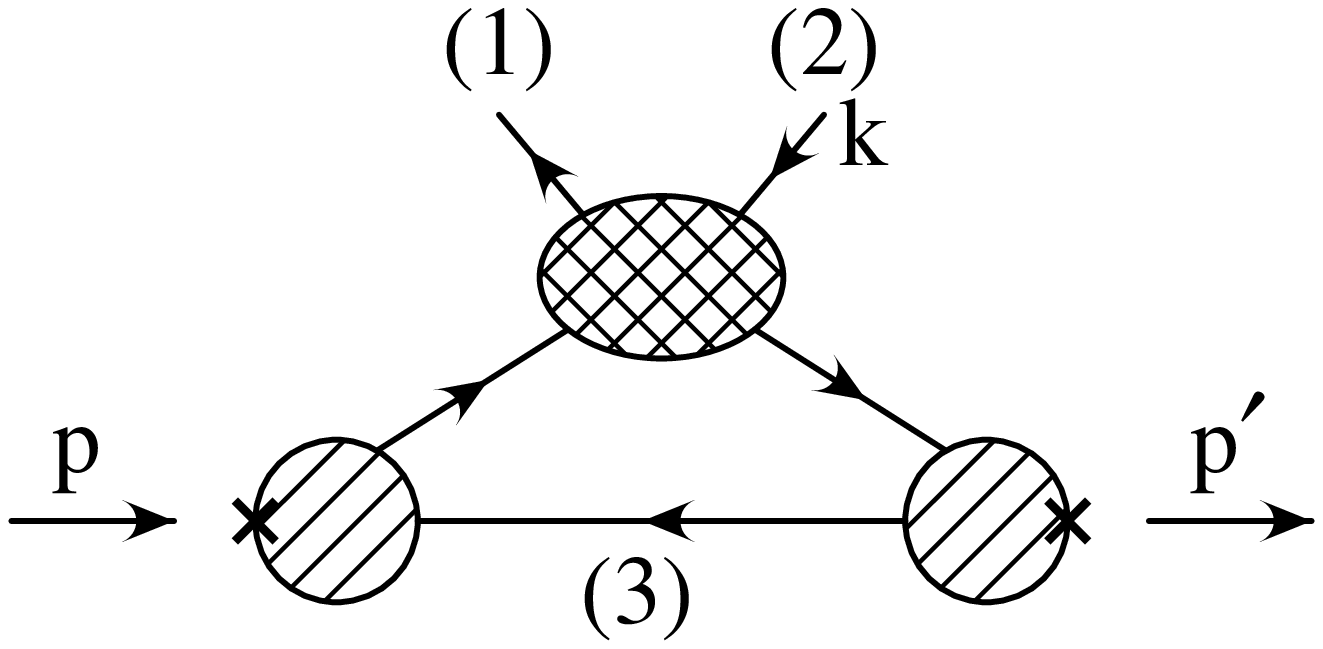}
\end{center}
\caption{\label{fig:softfn1} Non-resonant (left) and resonant (right) contribution to the Green's
function for the soft function. The crosses denote insertions of the
$b^\dagger_Rd_R^{\protect\phantom{\dagger}}=\psi_{1R}^\dagger
\psi_{3R}^{\protect\phantom{\dagger}}$ current on the left and the $d^\dagger_R
u_R^{\protect\phantom{\dagger}}=\psi^\dagger_{3R}\psi_{2R}^{\protect\phantom{\dagger}}$ current on the right. The outgoing
external fermion is a $b$-quark, the incoming external line is a
$u$-quark and the internal line is a $d$-quark.}
\end{figure}

Explicit expressions for the form factors were determined in
\cite{Grinstein:1994nx} by studying the crossed channel
(\ref{ffxdefd}). There it was shown that
\begin{align}
f_+(q^2)&=\sum_{\text{even $n$}}\frac{c_ng_{\pi Bn}(q^2)}{q^2-\mu_n^2}\\
\label{ffminus}
f_-(q^2)&= \sum_{\text{odd $n$}}\frac{c_ng_{\pi Bn}(q^2)}{q^2-\mu_n^2}
-\sum_{\text{even $n$}}\frac{c_ng_{\pi Bn}(q^2)(\mu_B^2-\mu_\pi^2)}{q^2(q^2-\mu_n^2)},
\end{align}
where the label $n$ refers to the tower of states with quantum numbers
$1\bar2=b\bar d$. The quantities $g_{\pi Bn}$, which are  interpreted as
triple meson couplings, are given by triple overlaps of
wave-functions. In terms of $\tilde\omega=p_-/q_-$, which is a function of
$q^2$ only,  it was found that, for even $n$
\begin{equation}
\label{gpiBneven}
g_{\pi Bn}(q^2)=\frac{-2\mu_n^2}{q^2\tilde\omega-\mu_B^2/ \tilde \omega}\,\hat g_n
\end{equation}
while for odd $n$
\begin{equation}
\label{gpiBnodd}
g_{\pi Bn}(q^2)=-2\mu_n^2\hat g_n
\end{equation}
where
\begin{equation}
\label{gngiven}
\hat g_n=\frac{1}{1-\tilde\omega}\int_0^{\tilde\omega}\!\dd z 
\phi^{(13)}(\frac{z}{\tilde\omega})\Phi^{(32)}(\frac{z-\tilde\omega}{1-\tilde\omega})
\phi_n(z)
-\frac{1}{\tilde\omega}\int_{\tilde\omega}^1\!\dd z 
\Phi^{(13)}(\frac{z}{\tilde\omega})\phi^{(32)}(\frac{z-\tilde\omega}{1-\tilde\omega})\phi_n(z)~.
\end{equation}

As shown in Ref.~\cite{Einhorn:1976uz} the form factors are
super-convergent: at large $q^2$
\beq
f_\pm \sim \frac1{|q^2|^{1+\beta_2}}.
\eeq
By considering the integral $\oint_C dz f_\pm(z)/(z-q^2)$ over a contour $C$
consisting of a circle at infinity deformed on the real line to avoid
all the poles, one can  show that the $q^2$ dependent couplings
may be replaced by their value on-shell, {\it e.g.,}
\beq
\label{eq:fplusonshell}
f_+(q^2)=\sum_{\text{even $n$}}\frac{c_ng_{\pi Bn}(q^2)}{q^2-\mu_n^2}
=\sum_{\text{even $n$}}\frac{c_ng_{\pi Bn}(\mu_n^2)}{q^2-\mu_n^2}
\eeq
This is useful because one can show\cite{Grinstein:1994nx} that in the
chiral limit, $m_2=m_3\to0+$, one has $g_{\pi Bn}(\mu_n^2)\to0$ for $n\neq0$, and
$g_{\pi BB}(\mu_B^2)\to -\mu_B^2/c_\pi $ where $c_\pi=c^{(23)}_0$ is the
normalized ``pion'' decay constant. Therefore, in the chiral limit
\beq
\label{eq:fpluschirallimit}
f_+(q^2)=-f_-(q^2)=\frac{f_B/f_\pi}{1-q^2/ \mu_B^2}
\eeq

\subsection{Soft Functions}
Next we turn our attention to the non-diagonal analog of the shape
function, the ``soft function'' defined, in analogy with the shape
function, by 
\beq
\label{eq:softdefd}
F_\mu\equiv\int \frac{\dd x^-}{4\pi}e^{-i\xi p_-x^-} 
\vev{\pi(p')|\bar u(x^-) P[e^{i\int_0^{x^-}\!\!\!\! A}]\gamma_\mu b(0)|B(p)}~.
\eeq
The soft function arises naturally in computations of $B$-meson decay
amplitudes in SCET in which a
pion is produced and is soft in the rest frame of the decaying
$B$-meson\cite{GPchiral,Grinstein:2005nu,Grinstein:2005ud}. 

The computation of the soft functions proceeds in complete analogy to
that of the previous section. 
The first contribution to (\ref{eq:softdefd}), shown in  Fig.~\ref{fig:softfn1}, is
similar to what we found for the shape function above in the region
$0<k_-<p'_-$. But now there is an
important difference, the first graph in Fig.~\ref{fig:softfn1} also gives a
non-vanishing  contribution from the region
$p'_--p_-<k_-<0$.  This new term cancels against a similar term from
the second graph in Fig~\ref{fig:softfn1}.  The result is
%
\begin{equation}
\label{softfnresult}
F_-=
 \phi^{(13)}(1+\xi-\omega)\phi^{(32)}(1-\xi/\omega)\theta(\omega-\xi)\theta(\xi)
+\sum_n\frac{\phi_n^{(12)}(1+\frac{\xi}{1-\omega}) g_n(q^2)}{q^2-\mu_n^2}
\theta(1+\xi-\omega)\theta(-\xi).
\end{equation}
We have introduced $\omega=p'_-/p_-$ and assumed $\omega<1$.  The meaning of $\xi$ is clear,
$\xi=k_-/p_-$. The functions $\phi^{(13)}$ and
$\phi^{(32)}$ correspond to the incoming and outgoing $B$ and $\pi$
mesons, respectively. The masses $\mu_n$ are for the tower of
states composed of $b$ and $\bar u$ quarks, and  $g_n(q^2)$ are
functions of $q^2=(p-p')^2$ given by
\begin{equation}
\label{gntilde}
g_n(q^2)=
\frac{1}{\omega} 
\int_\omega^1\!\dd{v}\,\phi^{(13)}(1-v)\Phi^{(32)}(v/\omega)\phi_n^{(12)}(\frac{1-v}{1-\omega})
-\frac1{1-\omega}\int_0^\omega
\!\dd{v}\,\phi^{(13)}(1-v)\phi^{(32)}(v/\omega)\Phi_n^{(12)}(\frac{1-v}{1-\omega})
\end{equation}

It is straightforward to check that this result gives correctly the
current form factor. Upon integration over $\xi$:
\begin{equation}
\label{ff1}
\vev{\pi(p')|\bar \psi^{(2)}\gamma_- \psi^{(1)}|B(p)}=p_-
\kappa(q^2)+ (p-p')_-\sum_n\frac{c_n  g_n(q^2)}{q^2-\mu_n^2}
\end{equation}
where $g_n$ was defined in (\ref{gntilde}) and  $\kappa$ is given by
\begin{equation}
\kappa(q^2)=\int_0^{\omega}\dd\xi \phi^{(13)}(1+\xi-\omega)\phi^{(32)}(1-\xi/\omega)~,
\end{equation}
in agreement with the classical result of \cite{Einhorn:1976uz}.

It will be useful to consider the crossed channel soft function. This
is interesting in it's own right, but more importantly, it will be
needed to investigate the chiral limit of the soft
functions. The graphs are again given by  Fig.~\ref{fig:softfn1}, 
except with the direction of the pion momentum reversed.
By direct computation we find
\begin{equation}
\label{softminusx}
F_-=-\sum_n\frac{\phi_n(x)\hat g_n(q^2)}{q^2-\mu_n^2}\, \theta(x)\theta(1-x)~,
\end{equation}
where $x$ is the momentum fraction of the outgoing light quark,
$x=1+\tilde\omega\xi=-k_-/(p_- +p'_-)$, and $\hat g_n(q^2)$ is given by
Eq.~(\ref{gngiven}). It is straightforward to check that this is in
fact the analytic continuation of the last term on the right hand side
of Eq.~(\ref{softfnresult}), that is, the soft function in the region
$-(p_--p_-')<k_-<0$.
 
We pause to make an interesting, though peripheral observation. The
soft function in Eq.~(\ref{softminusx}) vanishes for $k_- \geq 0$, while
the function in Eq.~(\ref{softfnresult}) is explicitly non-vanishing
in that region. Clearly one is not the analytic continuation (over
$q^2$ for fixed $k_-$) of the other. This violation of crossing
symmetry is not particularly worrisome. Crossing symmetry is not a
fundamental property of quantum field theories (and is not a
``symmetry''). It is established on a case by case basis (see, {\it
e.g.,} \cite{Weinberg:1995mt,Peskin:1995ev}).  It is surprising to see
it fail because one has learned from experience that without exception
one can show it. Except in the case at present. We suspect this case
does not satisfy crossing symmetry because we are not considering an
S-matrix element or a matrix element of a local operator but rather
the matrix element of a non-local operator partially Fourier
transformed. As soon as one integrates over $k_-$ crossing is
recovered: one obtains the form factors~(\ref{ffminus}) and~(\ref{ff1})
that can be easily checked to be the analytic continuation of each
other. Alternatively one may simply state that, in this case crossing
symmetry is the agreement between the analytic continuation of this
two expressions for $k_-<0$, and that crossing symmetry has nothing to
say about $k_- >0$.

The computation of $F_+$ is straightforward. Since $\bar \psi\gamma_+ \psi
=\psi^\dagger_L\psi_L^{\phantom{\dagger}}$, use of Eq.~(\ref{psiL}) indicates that the result is the
same as in (\ref{softminusx}) but with an additional factor of
$m_1m_2/4p_{1-}p_{2-}$, where $p_{1,2}$ are the momenta of the quarks
in Fig.~\ref{fig:softfn1}. Hence we obtain, in
the ``crossed'' channel, 
\begin{equation}
\label{softplusx}
F_+=\frac{m_1m_2}{2x(1-x)}\sum_n\frac{\phi_n(x)g_n(q^2)}{q^2-\mu_n^2}\,\theta(x)\theta(1-x)~,
\end{equation}
where $x$ is the momentum fraction of the outgoing light quark, 
$x=1+\tilde\omega\xi=-k_-/(p_- +p'_-)$. It is now easy to check that this
gives the current form factor $f_+$ upon integration over $x$. One
only needs to use Eq.~\eqref{eq:badphiintegral}
and the result, from Ref.~\cite{Grinstein:1994nx}, that
\beq
\sum_{n}\frac{c_ng_{\pi Bn}}{\mu^2_n}=0~.
\eeq

\subsection{Chiral Limit}
Since the form factors simplify tremendously in the chiral limit, it
is natural to ask whether this is also the case of the soft functions.
Moreover, the chiral limit is well understood in four dimensions, and
the methods used in understanding the chiral limit in four dimensions
should apply as well in two dimensions.

Consider the function 
\begin{equation}
\label{Gfunc}
G(x,q^2)=\sum_n\frac{\phi_n(x)\hat g_n(q^2)}{q^2-\mu_n^2},
\end{equation}
which appears in (\ref{softminusx}). We show in the Appendix
that for fixed $x$ this sum is super-convergent, that is
\begin{equation}
\label{Gfunc-scale}
G(x,q^2)\sim \frac1{(q^2)^{(1+\beta_3)}}~,\quad\text{as $q^2\to\infty $.}
\end{equation}
Hence one may integrate this in the complex $q^2$ plane over a closed
contour that avoids the positive real axis and closes on a circle at
infinity to show, just as for the form factor, that the numerators can
be replaced by residues:
\begin{equation}
G(x,q^2)=\sum_n\frac{\phi_n(x)\hat g_n(\mu_n^2)}{q^2-\mu_n^2},
\end{equation}

We can now use the result that in the chiral limit the couplings to
higher resonances vanish to simplify the soft functions of
Eqs.~\eqref{softminusx} and~\eqref{softplusx}, as follows:
\begin{align}
F_- &\to - \frac{\phi_B(x)\hat g_0(\mu_B^2)}{q^2-\mu_B^2}\,\theta(x)\theta(1-x)~,\\
F_+ &\to \frac{m_1m_2}{2x(1-x)} \frac{\phi_B(x)\hat g_0(\mu_B^2)}{q^2-\mu_B^2}\,\theta(x)\theta(1-x)~.
\end{align}
We have retained the vanishingly small  mass $m_2$ in the expression
for $F_+$ above. The chiral limit should be understood as having
arbitrarily small mass, but care must be exercised in taking the limit
$m_2=0$. For example, $m_2$ plays a role regulating the behavior
of $F_+$ as $x\to 0$. One can integrate over $x$ to obtain a form
factor, and then take the limit $m_2\to0$ smoothly.

To make this result completely explicit it is necessary to understand
the chiral limit of $\hat g_0(\mu_B^2)$. There is a subtlety here that
must be addressed with care. For small but nonzero ``pion'' mass,
$\mu_\pi=\mu^{(23)}_0$, the pole at $q^2=\mu_0^2$ falls bellow
threshold. Therefore the value of the variable $\tilde\omega$ at
$q^2=\mu_B^2$ is complex, and satisfies
\beq
\label{eq:tildeomegaonshell}
 (1-\tilde\omega)^2= -\frac{\mu_\pi^2}{\mu_B^2}+{\cal O}\left(\frac{\mu_\pi}{\mu_B}\right)^3
\eeq
Comparing the exact expression for the form factor $f_+$,
Eq.~\eqref{eq:fplusonshell}, with the result in the chiral limit,
Eq.~\eqref{eq:fpluschirallimit}, and using the explicit form of the
triple boson coupling given in Eqs.~\eqref{gpiBneven}--\eqref{gngiven},
we obtain the chiral limit for $\hat g_0$:
\beq
\label{eq:g0limit}
\hat g_0(\mu_B^2)\to -\frac{\mu_B^2}{c_\pi}(1-\tilde\omega)~.
\eeq
One can then check consistency by examining the chiral limit of the
second form factor, $f_-$. A similar computation yields
\beq
\hat g_0(\mu_B^2)\to \frac{\mu_\pi^2}{c_\pi (1-\tilde\omega)}~,
\eeq
which is seen to coincide with Eq.~\eqref{eq:g0limit} when use is made
of the resonance value of $\tilde\omega$ as given in
Eq.~\eqref{eq:tildeomegaonshell}. 

Collecting results and expressing them in terms of the light-cone
wave-functions, we see that, in the crossed channel, 
\begin{align}
F_- & \to
2\mu_B^2\frac{f_B}{f_\pi}\frac{p'_-}{p_-+p'_-}\frac{\tilde\phi_{B-}(x)}{q^2-\mu_B^2}~,\\
F_+ & \to
-2\mu_B^2\frac{f_B}{f_\pi}\frac{p'_+}{p_+}\frac{\tilde\phi_{B+}(x)}{q^2-\mu_B^2}~,
\end{align}

Finally, we can use these results to find the chiral limit of the soft
functions in the ``normal'' channel. Since $F_-$ in the crossed channel
as given by Eq.~\eqref{softminusx}  is the analytic
continuation of the second term in Eq.~(\ref{softfnresult}), we may
express the soft function in the ``normal'' channel  simply as
\begin{equation}
\label{softfnTMPresultchiral2}
-2\mu_B^2\frac{f_B}{f_\pi}\frac{p'_-}{p_--p'_-}\frac{\tilde\phi_{B-}(y)}{(p-p')^2-\mu_B^2}~,\\
\end{equation}
for $-(p_- - p'_-)< k_-<0$. Here  $y$ is  the argument of $\phi^{(12)}_n$
in \eqref{softfnresult}, namely, $y=1+\xi/(1-\omega)=(p_--p'_-
+k_-)/(p_--p_-')$, which is interpreted as the momentum fraction
carried by the $b$ quark. Moreover, for  $0<k_- <p'_-$ we had the result 
\begin{equation}
\label{softfnTMPresultchiral1}
\phi^{(13)}(1+\xi-\omega)\phi^{(32)}(1-\xi/\omega)
\end{equation}
which can be slightly simplified in the chiral limit, since the pion
wave-function simplifies in that limit, $\phi_\pi\to1$. Combining partial results we
have as our final result the chiral limit of the soft function in the
normal channel: 
\begin{equation}
\label{eq:Fmchiral}
F_-\to \frac{f_B}{f_\pi}\bigg[
\tilde \phi_{B-}((1-\omega)y)\theta(\xi)\theta(\omega-\xi)
-2\mu_B^2\frac{p'_-}{p_--p'_-}\frac{\tilde\phi_{B-}(y)}{(p-p')^2-\mu_B^2}\theta(-\xi)\theta(1+\xi-\omega)\bigg]
\end{equation}
Here, we remind the reader, $\omega=p'_-/p_-$, $\xi=k_-/p_-$, $y=1+\xi/(1-\omega)=(p_--p'_-
+k_-)/(p_--p_-')$ are the relevant momentum fractions 
 and we assumed $\omega<1$. 

\section{Discussion}
\label{sec:DISC}
\subsection{Shape Functions}
The result for the shape function, Eq.~\eqref{shfnresult}, is rather
surprising. The shape function ``factorizes'' into the product of
light-cone wave functions for the incoming and outgoing mesons, but
there is no reason to suspect this factorization a priori.  Certainly
na\"\i ve factorization would {\it not} have given this result. Na\"\i ve
factorization prescribes that one is to insert the vacuum in all
possible ways into the operator that is being sandwiched with meson
states, and extract the color singlet component only. But in this case
the operator is a non-local quark bilinear. If one were to insert the
vacuum between quark fields the resulting operator would have to be
set to zero by this prescription, since it has no color
singlet. Worse, vanishing of this quantity is protected also by
conservation of quark number. 

There is little guidance for phenomenological models of the shape
function. If one could only proof that the factorization observed in
this paper is a more general consequence of the large $N_c$ limit,
beyond the simple 1+1 dimensional case, it would be interesting to use
this as a first approximation to the shape function, with corrections
that vanish in the large $N_c$ limit. Alternatively, since the shape
function is directly measurable (up to power corrections) one could
use this relation to infer the light-cone wave-function, which
appears in many calculations but cannot be measured directly.  
Brave souls may pursue this
approximation as a conjecture, awaiting a proof in 4 dimensions:
\beq
\label{shape-result}
f_{B-}(k_-)=\kappa \left[\tilde\psi_{B-}(\bar\Lambda-k_-)\right]^2\theta(\bar\Lambda-k_-)
\eeq
where $f_B$ and $\tilde\psi_B$ stand for the shape and light-cone
wave-functions, $\bar\Lambda\equiv \Lambda_0=\mu_B-m_1$, and $\kappa^{-1}=\int_{-\bar\Lambda}^\infty \tilde\psi^2_{B-}(k_-)$. 

This relation may give interesting constraints on both functions since
both are separately constrained both by theory ({\it e.g.,} through
moments) and phenomenologically.  For example, consider popular
models for the light-cone wave-function  and shape function of
Refs.~\cite{Grozin:1996pq} and~\cite{Mannel:1994pm}, respectively:
\begin{gather}
\tilde\psi_{B-}^{\text{(model)}}=\frac{k_-}{K}\, \exp\left[-\frac{k_-}{K}\right],\\
f_{B-}^{\text{(model)}}=\frac{32}{\pi^2\bar\Lambda^3}(\bar\Lambda
-k_-)^2\exp\left[-\frac{4}{\pi\bar\Lambda^2}(\bar\Lambda -k_-)^2\right] \theta(\bar\Lambda
-k_-),
\end{gather}
where $K$ and $\bar\Lambda$ are positive constants. While clearly these functions fail
to satisfy the relation \eqref{shape-result}, the modifications needed
to bring them into compliance with \eqref{shape-result} are mild
(since both are exponentials times polynomials).  Whether this can be
done while still satisfying general guiding principles  in
Refs.~\cite{Grozin:1996pq} and~\cite{Mannel:1994pm} used to propose these
models is a question we do not embark on here.

\subsection{Soft function}
\label{subsec:SF}
It was shown in \cite{GPchiral} that the soft functions can be
expressed in terms of the light-cone wave-functions in the chiral
limit. Let us review the results from that paper. 

We begin by reviewing the effective Lagrangian for heavy mesons
interacting with low energy pseudo-Goldstone bosons, the so called
Heavy Meson Chiral Perturbation Theory (HMCHPT)\cite{wise,BuDo,TM}.
The effective Lagrangian, adapted to the two dimensional theory, is
written in terms of a heavy meson super-field,
\beq
H_a=-\left(\frac{1+\vslash}{2}\right)\gamma_5 B_a~,
\eeq
where $B_a=(b_u,b_d,B_s)$ are the positive energy fields for the $B$
mesons, and a matrix of pseudo-Goldstone bosons,
\begin{eqnarray}
M = \left(
\begin{array}{ccc}
 {\frac1{\sqrt2}}\pi^0 +
{\frac1{\sqrt6}}\eta &
\pi^+ & K^+ \\
\pi^-& -{\frac{1}{\sqrt2}}\pi^0 + {\frac1{\sqrt6}}\eta&K^0 \\
K^- &\bar K^0 &- {\frac{2}{\sqrt6}}\eta \\
\end{array}
\right)
\end{eqnarray}
We are considering the case of three light flavors, since this is what
is phenomenologically relevant. Our computations in previous
sections apply since the third light flavor does not contribute in any
of the processes we have considered (to leading order in $1/N_c$).
The matrix of pseudo-Goldstone bosons appears in the Lagrangian through $\xi = e^{iM/f}$ (and
$\Sigma=\xi^2$). Under $SU(3)_L\times SU(3)_R$ these transform as,
\begin{equation}
\label{eq:ch-trans}
\Sigma \to L \Sigma R^\dagger\,,\qquad \xi \to L \xi U^\dagger = U \xi R^\dagger
\qquad\text{and}\qquad
H_a^{\phantom{\dagger}} \to   H_b^{\phantom{\dagger}} \ U^\dagger_{ba} ~,
\end{equation}
where the  transformation $U$, defined through Eq.~\eqref{eq:ch-trans},
depends non-linearly on the pseudo-Goldstone bosons. In terms of these
fields, the effective Lagrangian is, to lowest order in a momentum
expansion and $1/m_b$,
\begin{multline}
\label{eq:CHLag}
{\cal L} =
-2i\, \text{Tr}\,[ \bar H^{(Q)a} v_{\mu} \partial^{\mu} H_a^{(Q)}]  
+ \frac{f^2}{ 8}\, \text{Tr} \left( \partial^{\mu} \Sigma \partial_{\mu} \Sigma^\dagger \right)
+\lambda_0 \text{Tr}\,\left[ m_q \Sigma + m_q \Sigma^\dagger \right]\\
+i \, \text{Tr}\,[\bar H^{(Q)a} H_b^{(Q)} v^{\mu} \left( \xi^\dagger\partial_{\mu} \xi + \xi \partial_{\mu}
  \xi^\dagger \right)_{ba}] 
-\frac{g}2 \,\text{Tr}\,[\bar H^a H_b \gamma_{\nu} \gamma_5\left(\xi^\dagger \partial^{\nu} \xi - \xi\partial^{\nu}\xi^\dagger \right)_{ba} ]
+ \cdots 
\end{multline}

The coupling constant $g$ determines the coupling of $B$ mesons to
pseudo-Goldstone bosons. The $\pi-B-B$ interaction term in the Lagrangian
is
\beq
-i\frac{g}{f}\,\epsilon_{\mu\nu}v^\nu  B_b^{\phantom{\dagger}}\partial^\mu M_{ba}^{\phantom{\dagger}}B^\dagger_a
\eeq
This is to be compared with the coupling $g_{\pi BB}$ which corresponds
to a  Lagrangian interaction term between three pseudo-scalar mesons
\beq
\hat g_{ijk}\epsilon_{\mu\nu}\partial^\mu\varphi_i\partial^\nu\varphi_j\varphi_k~.
\eeq
Here we have used the notation of Ref.~\cite{Grinstein:1994nx}. In
particular, we have a dimensionfull normalization constant relating
the coupling with a carat and the couplings in Eqs.~\eqref{gpiBneven}
and~\eqref{gpiBnodd}, $g_{\pi BB}=m_B^2 \hat g_{\pi BB}$. Setting
$\varphi_i(x)=\pi(x) $, $\varphi_j(x)=e^{-im_bv\cdot x}B(x)/ \sqrt{m_b}$ and
$\varphi_k(x)=e^{im_Bv\cdot x}B^\dagger(x)/ \sqrt{m_b}$ we see this is exactly the
same interaction as in the chiral Lagrangian with the identification
\beq
g_{\pi BB}=m_B^2 \hat g_{\pi BB}=m_B^2 \frac{g}{f}~.
\eeq
Below we use this relation together we the 't~Hooft model result $c_\pi \hat
g_{\pi BB}=-1$ to eliminate $g$ from the final expression for the
soft function; see Eq.~\eqref{eq:Fmchiraleff}.

Operators in the theory are also constrained by the symmetries. For
example, the left handed current $L^\nu_a=\bar q_a \gamma_\nu P_Lb$ can be
expressed in the low energy effective theory as\cite{wise,BuDo,TM}
\begin{equation}
L^\nu_a=-i\alpha \text{Tr}[\gamma^\nu P_L H_b^{\phantom{\dagger}} \xi^\dagger_{ba} ]~.
\end{equation}
The proportionality constant $\alpha$ can be fixed by requiring the
correct matrix element between a $B$-meson state and the vacuum,
$\alpha=\sqrt{m_B}f_B$ (we have adopted here non-relativistic normalization
of the $B$ states, as is standard practice in HQET).

\begin{figure}
\begin{center}
\includegraphics[height=3cm]{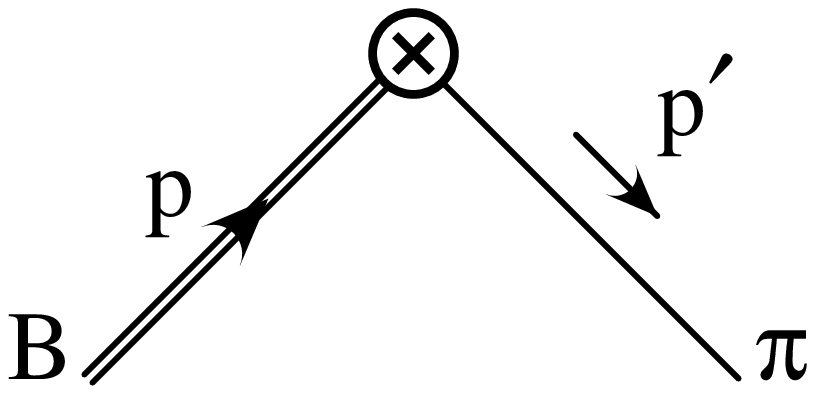}
\hspace{1in}
\includegraphics[height=3cm]{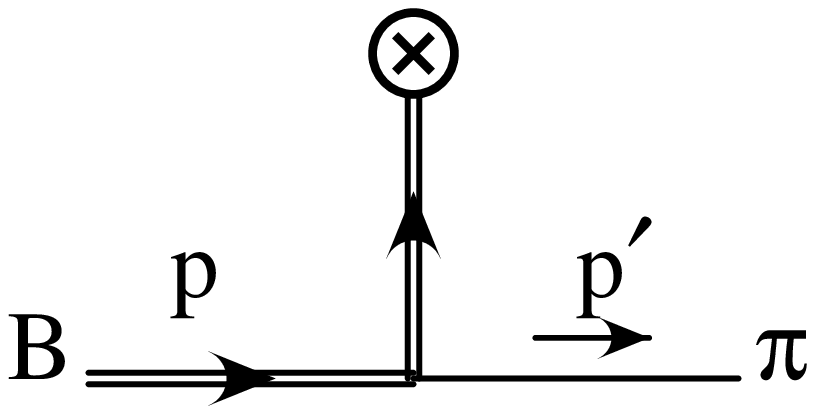}
\end{center}
\caption{\label{fig:chpt} Non-resonant (left) and resonant (right)
  contribution to  the soft function in Heavy Meson Chiral Perturbation Theory. 
The circled crosses denote insertions of the operators $O_{L,R}$ of
  Eq.~\protect\eqref{eq:OLRCHPTdefd}. }
\end{figure}

Following Ref.~\cite{GPchiral} we apply this technique to the
non-local operators in the definition of the soft functions. To follow
the notation in \cite{GPchiral} more closely, we
introduce the null vectors $n^\mu=(1/ \sqrt2,1/ \sqrt2)$ and $\bar
n^\mu=(1/ \sqrt2,-1/ \sqrt2)$. These have $n^+=\bar n^-=1$ and $n^-=\bar
n^+=0$. The non-local operators are
\begin{equation}
\label{eq:OLRdefd}
\begin{aligned}
O^\mu_{La}(k_-) &=\!\! \int \!\frac{\dd x^-\!\!}{2\pi}\, e^{-ix^-k_-}\text{T}\left[\bar
q_a(x^-)e^{i\int_0^{x^-} \!\!\!A}P_R \gamma^\mu b(0)\right],\\
O^\mu_{Ra}(k_-) &=\!\! \int \!\frac{\dd x^-\!\!}{2\pi}\, e^{-ix^-k_-}\text{T}\left[\bar
q_a(x^-)e^{i\int_0^{x^-}  \!\!\!A}P_L \gamma^\mu b(0)\right],
\end{aligned}
\end{equation}
which transform under the chiral group as $(\mathbf{\bar 3}_L, 1)$ and $(1,
\mathbf{\bar 3}_R)$, respectively. The corresponding operators in the effective
theory then are of the form
\begin{equation}
\label{eq:OLRCHPTdefd}
\begin{aligned}
O^\mu_{La}(k_-) &= \text{Tr}[\alpha_L(k_-)P_R\gamma^\mu H_b\,\xi^\dagger_{ba}]\\ 
O^\mu_{Ra}(k_-) &= \text{Tr}[\alpha_R(k_-)P_L\gamma^\mu H_b\,\xi_{ba}]
\end{aligned}
\end{equation}
Using the fact that $\vslash H = H$, the most general form of
$\alpha_{L,R}$ is a linear combination of $\nslash$ and
$\bnslash$. Matching to the matrix element between the vacuum and a
$B$ meson we obtain
\begin{equation}
\alpha_L(k_-)=\alpha_R(k_-) = -if_B/ \sqrt{m_B}[\nslash\tilde\psi_{B-}+\bnslash \tilde\psi_{B+}]~.
\end{equation}

It is now a straightforward exercise to compute the soft functions in
the effective theory. To compare with our earlier results for $F_-$ we
consider the sum of operators $O_L+ O_R$. There are two contributions,
as shown in Fig.~\ref{fig:chpt}: the first is a contact term obtained
by retaining a pion field in the expansion of $\xi_{ba}$ and the second
is a pole term from the no pion term in $\xi_{ba}$.
We obtain
\begin{equation}
\label{eq:Fmchiraleff}
i\sqrt{m_B}\vev{\pi(p')|(O_L+ O_R)_-|B(p)}=\frac{f_B}{f_\pi}\tilde\psi_{B-}(-k_-)
 \left[1+2\frac{\epsilon_{\nu\mu}v^\mu p^{\prime\nu}}{v\cdot(p-p')}\right].
\end{equation}
This is to be compared with our result, Eq.~\eqref{eq:Fmchiral}. For
the comparison we substitute $m_bv+p$ for $p$, so the meaning of $p$
becomes the residual momentum, and retain only the leading term in an
expansion in inverse powers of the heavy mass:
\begin{equation}
\label{eq:Fmchiral2}
F_-= \frac{f_B}{f_\pi}\bigg[
\tilde \psi_{B-}(p'_--k_-)\theta(k_-)\theta(p'_- -k_-)
-\frac{p'_-}{v_-}\tilde\psi_{B-}(-k_-)\theta(-k_-)\bigg]~.
\end{equation}
Using $v_-=v_+=1/ \sqrt2$ and neglecting $p'_+=2\mu_\pi^2/p'_-$ we see
that except for the argument of the light-cone wave-function and the
presence of $\theta$-functions the two terms in \eqref{eq:Fmchiraleff}
agree with the corresponding terms in \eqref{eq:Fmchiral2}.

The result is somewhat puzzling, since the form of
Eq.~\eqref{eq:Fmchiraleff} was obtained through general arguments in
Ref.~\cite{GPchiral}. However, the arguments presented there are not
quite rigorous. We can partially resolve the discrepancy as
follows. Consider gauging the vector flavor symmetries  of the QCD
Lagrangian. We gauge both the full, non-abelian, flavor symmetry of
the light quarks and the flavor symmetry of the heavy quark, ``$b$-number.''
The gauge fields are taken as background fields and are introduced
solely for the purpose of maintaining local
invariance, but will be set to zero at the end of the calculation. 
Now, instead of  the integrated operators in
\eqref{eq:OLRdefd}, we consider the
un-integrated, non-local operators
\begin{equation}
\label{eq:OLRdefd2}
\begin{aligned}
O^\mu_{La}(x^-,0) &= 
\text{T}(\bar q_a(x^-)e^{i\int_0^{x^-}\negthickspace\negthickspace\negthickspace A}P_R \gamma^\mu b(0),\\
O^\mu_{Ra}(x^-,0) &= 
\text{T}(\bar q_a(x^-)e^{i\int_0^{x^-}\negthickspace\negthickspace\negthickspace A}P_L \gamma^\mu b(0)~.
\end{aligned}
\end{equation}
Under a local vector flavor symmetry  transformation these transform
as
\beq
O^\mu_{L,R} \to B(0)O^\mu_{L,R}V(x^-)^\dagger ,
\eeq
where $B$ and $V$ stand for the unitary transformations of $b$-number
and light flavor, respectively. The corresponding operators
representing these in the chiral Lagrangian should transform the same
way. Hence we write
\begin{equation}
\label{eq:OLRCHPTdefd2} 
\begin{aligned}
O^\mu_{La}(x^-,0) &= \text{Tr}[\alpha_L(0,x^-)P_R\gamma^\mu
  H_c^{\phantom{\dagger}}(0)\beta_{cb}^{\phantom{\dagger}}(0,x^-)\xi^\dagger_{ba}(x^-)], \\
O^\mu_{Ra}(x^-,0) &= 
\text{Tr}[\alpha_R(0,x^-)P_L\gamma^\mu H_c^{\phantom{\dagger}}(0)\beta_{cb}^{\phantom{\dagger}}(0,x^-)\xi_{ba}^{\phantom{\dagger}}(x^-)].
\end{aligned}
\end{equation}
Here $\alpha_{L,R}$ are coefficient functions, to be determined, and
$\beta(0,x^-)=\text{T}\exp(-i\int_0^{x^-}A^V)$, where $A^V_\mu$ is the
background vector gauge field associated with the light quark flavor
vector symmetry, transforms as
\beq
\beta(0,x^-)\to V(0)\beta(0,x^-)V^\dagger(x^-).
\eeq

Turning off the background fields, Fourier transforming these
operators and repeating the argument above we now obtain
\begin{equation}
\label{eq:Fmchiraleff-new}
i\sqrt{m_B}\vev{\pi(p')|(O_L+ O_R)_-|B(p)}=
\frac{f_B}{f_\pi} \left[\tilde\psi_{B-}(p'_- -k_-)\theta(p'_- -k_-)
+2\tilde\psi_{B-}(-k_-)\theta(-k_-)\frac{\epsilon_{\nu\mu}v^\mu p^{\prime\nu}}{v\cdot(p-p')}\right]~.
\end{equation}
This only fails to reproduce Eq.~\eqref{eq:Fmchiral2} in that the
first term (the ``contact'' interaction) is missing the
$\theta$-function enforcing $k_- >0$. In the 't~Hooft model
calculation the origin of this restriction is clear: the contact terms
involves the product of the wave-functions of the $B$- and
$\pi$-mesons, and the latter requires $k_- >0$. However, the chiral
Lagrangian operator we have constructed has lost this information. It
would be interesting to see how this problem could be fixed. It seems
likely that some understanding of the representation of non-local
purely light-quark operators in the chiral/SCET Lagrangian is
necessary, since this should carry information about the pion
wave-function. However, since the physical basis for the restriction
that $k_- >0$ is pretty clear, it does make sense to adopt this result
in the full 4-dimensional analysis, now modified by replacing the
non-local operators \eqref{eq:OLRCHPTdefd2} for \eqref{eq:OLRCHPTdefd}.

\section{Conclusions}
\label{sec:CONC}
We have computed the analogues of the shape function and soft
functions of 4-dimensional HQET/SCET in QCD in 1+1 dimensions (the
``'t~Hooft Model''). These are matrix elements between mesons of
non-local operators. 

Our main results are as follows. First. the shape function, $f_B(k_-)$, is, up to a
multiplicative constant fixed by normalization conditions, the square
of the the light-cone wave function, $\tilde\psi_B(k_-)$,
\begin{equation*}
f_B(k_-)=\kappa \left[\tilde\psi_B(-k_-)\right]^2\theta(-k_-)~.
\end{equation*}
Second, in the chiral and heavy meson combined limits the soft
function for $B$ to $\pi$ transitions (defined in \eqref{eq:softdefd})
is completely determined by the $B$ meson light-cone wave-function, as
argued on general grounds  in
Ref.~\cite{GPchiral}. However, we find that the representation of the
non-local operator in the effective theory presented in
Ref.~\cite{GPchiral} is not correct. A better representation of this
operator is given by a non-local operator in the effective theory,
{\it e.g.,}
\begin{equation*}
O^\mu_{La}(x^-,0) = \text{Tr}[\alpha_L(0,x^-)P_R\gamma^\mu   
H_c^{\phantom{\dagger}}(0)\beta_{cb}^{\phantom{\dagger}}(0,x^-)\xi^\dagger_{ba}(x^-)]~.
\end{equation*}
This operator almost correctly reproduces the result from direct computation
in the 't~Hooft Model. It only fails to reproduce a $\theta$-function
restricting the momentum of the light quark (see Sect.~\ref{subsec:SF} for details),
which however is easily understood on physical grounds and can be
adopted (albeit in a somewhat ad-hoc fashion) in computations in four dimensional QCD. 

We cannot argue that the first of these results is applicable in four
dimensional QCD. However,  the two
results are tightly connected in the 't~Hooft model. This suggests
that there may in fact exist a way to justify the first result in four
dimensions. It is interesting to note in this regard that the same
argument that led us to consider non-local operators in the effective
theory for the soft-functions, gives that the effective operator in
the effective theory for the shape function is also non-local, roughly
\begin{equation}
O_\Gamma \sim \text{Tr}\,\left[\alpha(x^-,0)\bar H_a^{\phantom{\dagger}}(x^-)\beta(x^-,0)_{ab}^{\phantom{\dagger}} \Gamma H_b^{\phantom{\dagger}}(0)\right].
\end{equation}
Hence, if one could argue that $\alpha(x^-,0)$ factorizes, then the
result $f\propto\tilde\psi^2$ would follow automatically. Were this
established,  interesting phenomenological constraints
relating exclusive to inclusive $B$-decays would follow. Clearly this is a topic
that deserves much further investigation.

\begin{acknowledgments}
This work was suggested by Dan Pirjol, who, in addition, persuaded me
that there would be a modicum of interest in the results of this analysis.
This work was supported in part by the DOE under grant DE-FG03-97ER40546.
\end{acknowledgments}

\appendix*
\section{Scaling Theorems}
We follow the work of Einhorn closely\cite{Einhorn:1976uz}. Our scaling functions are
similar but not the same as his.
Let
\begin{equation}
\label{calGdefd}
{\cal G}^{(12)}(x,y;q^2)=\sum_n\frac{\phi_n^{(12)}(x)\phi_n^{(12)}(y)}{q^2-\mu_n^2}.
\end{equation}
One then shows this has a limit, which  defines the scaling functions
\begin{equation}
h_\pm^{(1)}(\xi;y)=\lim_{q^2\to\infty} q^2{\cal G}^{(12)}(\pm\frac{\xi}{q^2},y;q^2)
\end{equation}
Note that integrating $h(\xi;y)$ over $y$ gives Einhorn's scaling
function $h(\xi)$. 
The scaling function satisfies 
\begin{equation}
\left(1-\frac{\tilde m_1^2}{\xi}\right)h_\pm^{(1)}(\xi;y)
+\pvint_0^1\dd\eta \,\frac{h_\pm^{(1)}(\eta ;y)}{(\eta-\xi)^2}=\delta(y)~,
\end{equation}
from which it immediately follows that
\begin{equation}
h_\pm^{(1)}(\xi;y)\sim\xi^{\beta_1} \quad\text{as $\xi\to0$, fixed $y$.}
\end{equation}

Knowing that ${\cal G}$ has a finite limit one can determine the large $q^2$ scaling 
of $G(x,q^2)$ in Eq.~(\ref{Gfunc-scale}). To this end write
explicitly the first term in $g_n(q^2)$ from Eq.~(\ref{gngiven}) in
$G(x,q^2)$ as
\begin{multline}
\label{app1}
\frac{1}{1-\tilde \omega}\int_0^{\tilde\omega} \!\!\!\!\dd z
\phi^{(13)}(\frac{z}{\tilde\omega})
\Phi^{(32)}(\frac{z-\tilde\omega}{1-\tilde\omega})
\sum_n\frac{\phi_n^{(12)}(x)\phi_n^{(12)}(z)}{q^2-\mu_n^2}
\\
=\frac{\tilde\omega}{1-\tilde \omega}\int_0^{1} \!\!\!\!\dd v \int_0^1\!\!\!\!\dd t\,
\frac{\phi^{(13)}(v) \phi^{(32)}(t){\cal G}^{(12)}(\tilde\omega v,x;q^2)}{\left(
t+\frac{\tilde\omega}{1-\tilde\omega}(1-v)\right)^2}
\\
=\int_0^{1} \!\!\!\!\dd v \int_1^{1/ \tilde\omega}\!\!\!\!\dd u\,
\frac{\phi^{(13)}(v) \phi^{(32)}(\frac{\tilde\omega (u-1)}{1-\tilde\omega }){\cal G}^{(12)}(\tilde\omega v,x;q^2)}{(u-v)^2}
\end{multline}
In the second line we have changed integration variables ($v= z/
\tilde\omega$), used the integral representation of $\Phi$, and expressed the
sum in terms of the function ${\cal G}$ of (\ref{calGdefd}). The third
line uses a further change of variables $\tilde\omega
u=t(1-\tilde\omega)+\tilde\omega$ so
the denominator factors,
\begin{equation}
\big((1-\tilde\omega)t+\tilde\omega(1-v)\big)^2=\tilde\omega^2(u-v)^2,
\end{equation}
which leaves the integrand in a form suitable for exploration of the
small $\tilde\omega$ behavior. This corresponds to large $q^2$,
$q^2\sim m_B^2/\tilde\omega$. 

Using $\phi^{(32)}(x)\approx cx^{\beta_3}$ as $x\to0$, the behavior of (\ref{app1})
is obtained by writing it in terms of the scaling function:
\begin{equation}
\sim \frac{c\,(m_B^2)^{\beta_3}}{(q^2)^{(1+\beta_3)} }
\int_0^{1} \!\!\!\!\dd v \int_1^\infty\!\!\!\!\dd u\,
\frac{\phi^{(13)}(v)  (u-1)^{\beta_3}h_+^{(12)}(m_B^2 v;x)}{(u-v)^2}~.
\end{equation}

Analogous steps are taken for the second term in $G(x;q^2)$:
\begin{multline}
\label{app2}
-\frac{1}{\tilde \omega}\int_{\tilde\omega}^1 \!\!\!\!\dd z
\Phi^{(13)}(\frac{z}{\tilde\omega})
\phi^{(32)}(\frac{z-\tilde\omega}{1-\tilde\omega})
\sum_n\frac{\phi_n^{(12)}(x)\phi_n^{(12)}(z)}{q^2-\mu_n^2}
\\
=-\frac{1-\tilde \omega}{\tilde\omega}\int_0^{1} \!\!\!\!\dd v \int_0^1\!\!\!\!\dd t\,
\frac{\phi^{(13)}(t) \phi^{(32)}(v){\cal G}^{(12)}(\tilde\omega+(1-\tilde\omega) v,x;q^2)}{\left(
t-1-\frac{1-\tilde\omega}{\tilde\omega}\;v\right)^2}
\\
=-\int_0^{1} \!\!\!\!\dd t \int_1^{1/ \tilde\omega} \!\!\!\!\dd u\,
\frac{\phi^{(13)}(t) \phi^{(32)}(\frac{\tilde\omega(u-1)}{1-\tilde\omega})
{\cal G}^{(12)}(\tilde\omega u,x;q^2)}{(u-t)^2}
\\
\sim-\frac{c\,(m_B^2)^{\beta_3}}{(q^2)^{(1+\beta_3)} }
\int_0^{1} \!\!\!\!\dd t \int_1^\infty\!\!\!\!\dd u\,
\frac{\phi^{(13)}(t)  (u-1)^{\beta_3}h_+^{(12)}(m_B^2 u;x)}{(u-t)^2}~.
\end{multline}


\end{document}